\documentclass[english,letterpaper,aps,prl,twocolumn,showpacs]{revtex4-1}
\usepackage{lmodern}

\usepackage[T1]{fontenc}
\usepackage[latin1]{inputenc}
\usepackage{graphicx}
\usepackage{amssymb}

\makeatletter

\usepackage{lmodern}
\usepackage{subfig}

\makeatletter

\usepackage{lmodern}

\makeatletter

\usepackage{lmodern}

\makeatletter

\usepackage{lmodern}

\makeatletter

\makeatletter

\usepackage{epsfig}

\usepackage{dcolumn}

\usepackage{bm}

\usepackage{bbm}

\usepackage{wasysym}

\newlength{\textwidthm}

\makeatother

\makeatother

\makeatother

\makeatother

\makeatother

\usepackage{babel}
\makeatother
\begin{document}

\title{While calculating RKKY interaction in graphene no theorist should do a cut-off  without cause}

\author{E. Kogan}
\email{Eugene.Kogan@biu.ac.il}

\affiliation{Department of Physics, Bar-Ilan University, Ramat-Gan 52900,
Israel}
\date{\today}

\begin{abstract}
In our previous work (E. Kogan, Phys. Rev. B {\bf 84}, 115119 (2011)) we presented
 calculation of RKKY interaction between two magnetic impurities in graphene based on  Matsubara Green's functions (MGF) in the coordinate -- imaginary time  representation. Now we present the calculation based on   MGF in the  coordinate -- frequency representation.
We claim that both approaches have an important advantage over those based on zero temperature Green's functions (ZTGF), which are  very briefly reviewed  in the beginning of the present work. The MGF approaches, in distinction to the ZTGF approaches,  operate only with the convergent integrals from the start to the end of the calculation.
The  coordinate -- frequency representation for the MGF turns out to be as convenient as  the coordinate -- imaginary time  representation and allows to easily consider the cases of doped and gapped graphene.
\end{abstract}


\maketitle

\section{Introduction}

RKKY interaction between two magnetic impurities in graphene was theoretically studied quite intensely during last several years
 \cite{vozmediano,dugaev,brey,saremi,cheianov,black,sherafati,uchoa,bunder,power,kogan,sherafat,szalowski,bunder2,kettemann,sherafati2,liwei,loss}. (A terse but precise review
 of the issue one can find in the book by M. Katsnelson \cite{katsnelson}.)
One may ask, why  the problem, which is  in principle so simple (when being treated in the lowest order of perturbation theory,
like it was done in all the papers referenced above), was the subject of so many publication, using different approaches?
The answer to this question, as presented below, is connected with the fact that a simply written integral is not necessarily a simply calculated  integral,
and in the frame work of all zero temperature Green's functions (ZTGF) the integrals defining the RKKY interaction in graphene   turned out to be divergent. Different ZTGF approaches can be thus viewed as different ways to obtain  finite results from the divergent integrals.

In  our previous publication dealing with the subject we were using the approach
based on  Matsubara Green's functions (MGF) in the coordinate -- imaginary time  representation \cite{kogan}. In this work we consider RKKY interaction in graphene in the framework of approach based on  MGF in the  coordinate -- frequency representation, and find that this approach is no less convenient than the one we used previously. Before we present the approach, we give
a brief review of the existing ZTGF approaches, indicating explicitly where  the divergent integrals appear.

If we ignore the spin--orbit coupling,
the effective exchange  RKKY  interaction between the two magnetic impurities with the  spins ${\bf S}_1$ and ${\bf S}_2$,   sitting on top of carbon atoms at the sites  $i$ and $j$, is
\begin{eqnarray}
\label{abr5}
H_{RKKY}=-\frac{1}{4}J^2\chi(i,j){\bf S}_1{\bf\cdot  S}_2,
\end{eqnarray}
where $J$ is the contact exchange interaction between each of spins and the graphene electrons, and $\chi_{R}$ is the free electrons charge  susceptibility.
Thus  different approached to  calculation of the RKKY interaction are actually different approaches to calculation the susceptibility.

Not to distract attention of the reader from the aspects of the physics we are going to concentrate upon,
we'll consider a toy model of graphene, with free electrons being described by the 2d Dirac Hamiltonian. Thus the model can possess only a single Dirac point,
and we'll present the existing approaches as if they were applied to this toy model. Notice, that due to the isotropy of the model
\begin{eqnarray}
\chi(i,j)=\chi(R),
\end{eqnarray}
where $R$ is the distance between the sites $i$ and $j$.

\section{Approaches based on ZTGF}

The  approach, used  in Refs. \cite{vozmediano,brey,saremi} is based on equation
\begin{eqnarray}
\label{abr9}
\chi(R)=\int \frac{d^2{\bf q}}{(2\pi)^2}\chi(\omega=0,{\bf q})e^{i{\bf q\cdot R}},
\end{eqnarray}
where
\begin{eqnarray}
\chi(\omega=0,{\bf q})=2\int\frac{d^2{\bf k}}{(2\pi)^2}\frac{n_F(\xi_{\bf k})-n_F(\xi_{{\bf k}+{\bf q}})}{E_{{\bf k}+{\bf q}}-E_{\bf k}};
\end{eqnarray}
$\xi_n=E_n-\mu$ and  $n_F(\xi)=\left(e^{\beta\xi}+1\right)^{-1}$ is the Fermi distribution function.
This approach, though looking quite straightforward, brings with it a problem. In  a model of infinite Dirac cones for
$\chi(\omega=0,{\bf q})$ we obtain a diverging integral. To obtain  finite values from these divergent integrals, as it was mentioned previously,  one has
to implement the complicated (and to some extent arbitrary) cut-off procedure \cite{saremi}.

The problem can be formulated in a different way. Being calculated in a  realistic band model, with the bands of finite width, $\chi(\omega=0,{\bf q})$
is not a universal quantity. It depends not only on infrared physics, but on the properties of electron spectrum and eigenfunctions in the whole Brillouin zone (even for small ${\bf q}$).

Another approach, formulated in Ref. \cite{dugaev}, starts from a well known equation for the susceptibility
\begin{eqnarray}
\label{dugaev}
\chi(R)= \frac{2i}{\pi}\int_{-\infty}^{\infty}G^2(R,E)dE,
\end{eqnarray}
where $G$ is the retarded green's function. Here again  the integral diverges on both limits of integration. However the authors changed the contour of integration, transforming the divergent integral (\ref{dugaev}) into the convergent integral  along the imaginary axis (see also Ref. \cite{duga}).   The authors also considered RKKY interaction in  gapped graphene, when the  power law  decrease of the  interaction with the distance turns into the exponential law. Actually, in a implicit form the authors made the transition from ZTGF to MGF, so our final results will be very close to ones obtained  in Ref. \cite{dugaev}.

The approach, using formula
\begin{eqnarray}
\chi(r,r')=\delta n(r)/\delta V(r')
\end{eqnarray}
and, hence, calculating electron susceptibility on the basis of equation
\begin{eqnarray}
\label{sherafati}
\chi(R)= -\frac{2}{\pi}\int_{-\infty}^{E_F}\text{Im}\left[G^2(R,E)\right]dE,
\end{eqnarray}
where  $E_F$ is the Fermi energy, was first used, in application to graphene to the best of our knowledge, in Ref. \cite{sherafati}.
An advantage of this approach is that it allows to easily consider the case of doped graphene, the disadvantage is that the approach, like  the one presented above, has to deal with the divergent integral (the integral with respect to $dE$ diverges at the lower limit of integration). Also in this case,
to obtain  finite values from these divergent integrals  one has
to implement the complicated (and to some extent arbitrary) cut-off procedure.

\section{MGF in frequency representation}

Our approach will be based on equation \cite{abrikosov}
\begin{eqnarray}
\label{abr99}
\chi(R)=-\frac{1}{\pi}\int_{-\infty}^{\infty}{\cal G}^2(R;\omega)d\omega,
\end{eqnarray}
where ${\cal G}(R;\omega)$ is the MGF in frequency momentum representation.
Dirac equation describing electrons is
\begin{eqnarray}
\label{dirac}
H=v(\tau^xk_x+\tau^yk_y),
\end{eqnarray}
where the matrix ${\bf \tau}$ acts in the space of two sublattices. From Eq. (\ref{dirac}) we obtain
\begin{eqnarray}
\label{matsu}
{\cal G}(k,\omega)=\frac{1}{i\omega+\mu-v(\tau^xk_x+\tau^yk_y)},
\end{eqnarray}
where $\mu$ is a chemical potential.
From Eq. (\ref{matsu}) we obtain \cite{shytov}
\begin{eqnarray}
&&{\cal G}^{CC}(R,\omega)=\frac{-i(\omega-i\mu)}{(2\pi)^2}\int_0^\infty\frac{kdk}{(\omega-i\mu)^2+v^2k^2} \nonumber\\
&&\cdot\int_0^{2\pi}e^{ikR\cos\theta}d\theta=\frac{-i(\omega-i\mu)}{2\pi}\int_0^\infty\frac{kJ_0(kR)dk}{(\omega-i\mu)^2+v^2k^2} \nonumber\\
&&=\frac{-i(\omega-i\mu)}{2\pi v^2}K_0\left[\text{sign}(\omega)(\omega-i\mu)R/v\right].
\end{eqnarray}
\begin{eqnarray}
&&{\cal G}^{AB}(R,\omega)=\frac{-v}{(2\pi)^2}\int_0^\infty\frac{k^2dk}{(\omega-i\mu)^2+v^2k^2} \nonumber\\
&&\cdot\int_0^{2\pi}e^{ikR\cos\theta+i\theta}d\theta=\frac{-v}{2\pi}\int_0^\infty\frac{k^2J_1(kR)dk}{(\omega-i\mu)^2+v^2k^2} \nonumber\\
&&=\frac{-\text{sign}(\omega)(\omega-i\mu)}{2\pi v^2}K_1\left[\text{sign}(\omega)(\omega-i\mu)R/v\right].
\end{eqnarray}
($CC$ can mean either $AA$ or $BB$; $K_0$ and $K_1$ are the modified Bessel function of zero and first order respectively).
We have used mathematical identity, valid for Re $z>0$ \cite{prudnikov},
\begin{eqnarray}
\int_0^{\infty}\frac{x^{\nu+1}}{(x^2+z^2)^{\rho}}J_{\nu}(cx)dx=\frac{c^{\rho-1}z^{\nu-\rho+1}}{2^{\rho-1}\Gamma(\rho)}K_{\nu-\rho+1}(cz).\nonumber\\
\end{eqnarray}

\section{Undoped graphene}

Consider first the case of undoped graphene ($\mu=0$).

\subsection{MGF in frequency representation}

Using another  mathematical identity \cite{prudnikov}
\begin{eqnarray}
&&\int_0^{\infty}x^{\alpha-1}K_{\mu}(cx)K_{\nu}(cx)dx=\frac{2^{\alpha-3}}{c^{\alpha}\Gamma(\alpha)}\Gamma\left(\frac{\alpha+\mu+\nu}{2}\right)\nonumber\\
&&\Gamma\left(\frac{\alpha+\mu-\nu}{2}\right)\Gamma\left(\frac{\alpha-\mu+\nu}{2}\right)\Gamma\left(\frac{\alpha-\mu-\nu}{2}\right),
\end{eqnarray}
from Eq. (\ref{abr99}) we obtain \cite{saremi}
\begin{eqnarray}
\label{final}
\chi_{\mu=0}^{AA}(R)=\frac{1}{64\pi vR^3},\qquad \chi^{AB}_{\mu=0}(R)=-\frac{3}{64\pi vR^3}.
\end{eqnarray}

\subsection{MGF in time representation}

For completeness, we reproduce here the calculation of $\chi$   based on MGF in coordinate -- imaginary time representation, presented in our previous paper \cite{kogan}. The susceptibility was written as  \cite{saremi,cheianov,kogan}
\begin{eqnarray}
\label{abr}
\chi(R)=-2\int_{-\infty}^{\infty}{\cal G}(R;\tau){\cal G}(R;-\tau)d\tau.
\end{eqnarray}
Transition from frequency to imaginary time representation yields the MGF
\begin{eqnarray}
{\cal G}^{CC}({\bf k},\tau)&=&\frac{\text{sign}(\tau)}{2} e^{-vk|\tau|}\nonumber\\
 {\cal G}^{AB}({\bf k},\tau)&=&\frac{1}{2} e^{-vk|\tau|+i\theta}.
\end{eqnarray}

As a result we obtain  \cite{kogan}
\begin{eqnarray}
\label{a}
{\cal G}^{CC}(R,\tau)&=&\frac{\text{sign}(\tau)}{8\pi^2}\int_0^{\infty}dk k\int_0^{2\pi} d\theta e^{ikR\cos\theta-vk|\tau|}\nonumber\\
{\cal G}^{AB}(R,\tau)&=&\frac{1}{8\pi^2}\int_0^{\infty}dk k\int_0^{2\pi} d\theta e^{ikR\cos\theta+ i\theta-vk|\tau|}.\nonumber\\
\end{eqnarray}
Performing   the angular integrations in Eq. (\ref{a})
we get
\begin{eqnarray}
\label{aaa4}
{\cal G}^{CC}(R;\tau)&=&\frac{\text{sign}(\tau)}{4\pi}\int_0^{\infty}dk kJ_0(kR)e^{-vk|\tau|}\nonumber\\
{\cal G}^{AB}(R;\tau)&=&\frac{1}{4\pi}\int_0^{\infty}dk kJ_1(kR)e^{-vk|\tau|}.
\end{eqnarray}
($J_0$ and $J_1$ are the Bessel function of zero and first order respectively).
Using mathematical identity \cite{prudnikov}
\begin{eqnarray}
\label{identity}
&&\int _0^{\infty}x^{n-1}e^{-px}J_{\nu}(cx)dx\\
&&=(-1)^{n-1}c^{-\nu}\frac{\partial^{n-1}}{\partial p^{n-1}}\frac{\left(\sqrt{p^2+c^2}-p\right)^{\nu}}
{\sqrt{p^2+c^2}},\nonumber
\end{eqnarray}
integrals in the RHS of Eq. (\ref{aaa4}) can be calculated exactly,
giving a well known result \cite{cheianov}
\begin{eqnarray}
\label{gg}
{\cal G}^{CC}(R;\tau)=\frac{1}{4\pi}\frac{v\tau }{(v^2\tau^2+R^2)^{3/2}}\nonumber\\
{\cal G}^{AB}(R;\tau)=\frac{1}{4\pi}\frac{R}{(v^2\tau^2+R^2)^{3/2}}.
\end{eqnarray}
The remaining integration in Eq. (\ref{abr}) is trivial; as a result we  recover
 Eq. (\ref{final}).

\section{Doped graphene}

In the case of doped graphene the susceptibility (\ref{abr99})  will be expressed through  the integrals
\begin{eqnarray}
\text{Re}\left\{\int_{0}^{\infty}(\omega-i\mu)^2K_{0,1}^2\left[(\omega-i\mu)R/v\right]d\omega\right\}.
\end{eqnarray}
Considering integrals in the complex plane it is convenient to deform the contour of integration and present the integrals as
\begin{eqnarray}
\text{Re}\left\{\int_{0}^{\infty}\omega^2K_{0,1}^2(\omega R/v)d\omega+\int^{0}_{-i\mu}\omega^2K_{0,1}^2(\omega R/v)d\omega\right\}.\nonumber\\
\end{eqnarray}
Taking into account the identity
\begin{eqnarray}
K_{\alpha}(-ix)=\frac{\pi}{2}i^{\alpha+1}[J_{\alpha}(x)+iY_{\alpha}(x)],
\end{eqnarray}
we  get \cite{sherafat}
\begin{eqnarray}
\label{mejer}
\chi_{\mu}^{CC}(R)=\chi_{\mu=0}^{CC}(R)\left[1-16\int_{0}^{k_FR}dzz^2J_{0}(z)Y_{0}(z)\right]\nonumber\\
\chi_{\mu}^{AB}(R)=\chi_{\mu=0}^{AB}(R)\left[1+\frac{16}{3}\int_{0}^{k_FR}dzz^2J_{1}(z)Y_{1}(z)\right],\nonumber\\
\end{eqnarray}
where $k_F=\mu/v$.
The integrals in Eq. (\ref{mejer}) can be presented in terms of Meijer functions \cite{sherafat} (I send the reader to
that Reference for the details).

It is interesting to compare the RKKY exchange in doped graphene, with its two sublattices and linear dispersion law,  with
that in ordinary two-dimensional electron gas. For the latter the Green's function is
\begin{eqnarray}
{\cal G}(k,\omega)=\frac{1}{i\omega+\mu-k^2/2m}.
\end{eqnarray}
Hence the susceptibility turns out to be  \cite{fischer}
\begin{eqnarray}
\chi(R)\sim\frac{1}{R^2}\int_{0}^{k_FR}dzzJ_{0}(z)Y_{0}(z).
\end{eqnarray}
Amusing, that the authors of Ref. \cite{fischer} were affiliated with the same University, as the author of the present paper.

\section{Gapped graphene}

Consider now graphene with the gap in electron spectrum described by Dirac Hamiltonian
\begin{eqnarray}
\label{dirac2}
H=v(\tau^xk_x+\tau^yk_y)+\Delta\tau^z.
\end{eqnarray}
The Green's function is
\begin{eqnarray}
\label{dirac26}
{\cal G}({\bf k},\omega)=\frac{-i\omega-\Delta\tau^z-v(\tau^xk_x+\tau^yk_y)}{\omega^2+\Delta^2+v^2k^2}.
\end{eqnarray}

\subsection{MGF in frequency representation}

From Eq. (\ref{dirac26}) we obtain
\begin{eqnarray}
\label{cc}
&&{\cal G}^{CC}(R,\omega)=\frac{-i\omega\mp\Delta}{2\pi}\int_0^\infty\frac{kJ_0(kR)dk}{\omega^2+\Delta^2+v^2k^2}\nonumber\\
&&=\frac{-i\omega\pm\Delta}{2\pi v^2}K_0\left(\sqrt{\omega^2+\Delta^2} R/v\right).
\end{eqnarray}
\begin{eqnarray}
\label{gap}
&&{\cal G}^{AB}(R,\omega)=\frac{-v}{2\pi}\int_0^\infty\frac{k^2J_1(kR)dk}{\omega^2+\Delta^2+v^2k^2}\nonumber\\
&&=\frac{-\sqrt{\omega^2+\Delta^2}}{2\pi v^2}K_1\left(\sqrt{\omega^2+\Delta^2} R/v\right).
\end{eqnarray}
In Eq. (\ref{cc}) minus corresponds to $AA$ and plus to $BB$. Substituting into Eq. (\ref{abr99}) we obtain \cite{dug}
\begin{eqnarray}
\label{gap3}
&&\chi^{CC}(R)=\frac{1}{4\pi^3 v^4}\nonumber\\
&&\int_{-\infty}^{\infty}(\omega^2-\Delta^2)K_0^2\left(\sqrt{\omega^2+\Delta^2} R/v\right)d\omega.
\end{eqnarray}
\begin{eqnarray}
\label{gap2}
&&\chi^{AB}(R)=-\frac{1}{4\pi^3 v^4}\nonumber\\
&&\int_{-\infty}^{\infty}(\omega^2+\Delta^2)K_1^2\left(\sqrt{\omega^2+\Delta^2} R/v\right)d\omega.
\end{eqnarray}

The remaining integrations can be  performed analytically in two limiting cases.

Consider first the case $\Delta R/v\ll 1$. Here it is appropriate to mention the relation between
the toy model, we are using, and real graphene. The existence of two Dirac points in graphene leads to additional  angular dependent factor in the formula for the RKKY interaction. It was thoroughly studied previously and does not interfere with the physics we are discussing in this work.
More interesting is the condition of the applicability of the infinite Dirac cones dispersion law to the calculation of RKKY interaction in graphene.
From Eq. (\ref{gg}) we see that characteristic values of $\tau$ entering into integral (\ref{abr}) are of the order of $R/v$; hence characteristic $k$ entering into integrals (\ref{aaa4}) are of the order of $1/R$. Thus the approximation of linear dispersion law   is applicable, provided $R\gg a$, where $a$ is the graphene lattice constant. Now we realize, that the case $\Delta R/v\ll 1$ can be described in the framework of the model
if $\Delta\ll av$, or in simple terms, if the gap is  narrow in comparison with  the graphene band width, which is certainly true in most cases of gapped  graphene. So finally, in the case of a narrow gap and relatively small distances we can go to the limit $\Delta\to 0$  in Eqs. (\ref{gap3}) and (\ref{gap2}),   and recover the results of the gapless case.

The other limiting case $\Delta R/v\gg 1$ is more interesting. In this case we may use asymptotic expression for  modified Bessel functions
\begin{eqnarray}
K_{\nu}(z)\sim\sqrt{\frac{\pi}{2z}}e^{-z}.
\end{eqnarray}
After calculating the resulting integrals in Eq. (\ref{gap2}) using the Laplace method, we obtain the same result both for inter-sublattice and intra-sublattice susceptibility
\begin{eqnarray}
\label{bu}
\chi=-\frac{1}{8v}\left(\frac{\Delta}{\pi vR}\right)^{3/2}e^{-2R\Delta/v}.
\end{eqnarray}

It is worth paying attention to the fact that Eq. (\ref{bu}) seems to contradict  rigorously proved theorem stating
that for any bipartite lattice at half filling, the RKKY interaction is antiferromagnetic
between impurities sitting on top of atoms belonging to opposite  sublattices (i.e., $A$ and $B$ sublattices in graphene), and is ferromagnetic between impurities
sitting on top of atoms
belonging to the same sublattice \cite{saremi,pereira,kogan}. However, the theorem is not applicable to Hamiltonian (\ref{dirac2}), with its last term meaning that if we rewrite  the Hamiltonian in the tight-binding representation, the
intra-sublattice hopping will appear, hence the lattice is no longer bipartite. More specifically, the spectrum still has the symmetry of that in bipartite lattice, but the wave functions do not \cite{kogan}. 

\subsection{MGF in time representation}

Here we'll restrict ourselves with the calculation of $\chi^{AB}$.
Transition from frequency to imaginary time representation yields the MGF
\begin{eqnarray}
\label{cc2}
{\cal G}^{AB}({\bf k},\tau)=\frac{1}{2}\frac{vk}{\sqrt{\Delta^2+v^2k^2}} e^{-\sqrt{\Delta^2+v^2k^2}|\tau|+i\theta}.
\end{eqnarray}
Hence instead of Eq. (\ref{aaa4}) we obtain
\begin{eqnarray}
&&J_0(kR)e^{-\sqrt{\Delta^2+v^2k^2}|\tau|}\\
&&{\cal G}^{AB}(R;\tau)=\frac{v}{4\pi}\int_0^{\infty}dk \frac{k^2}{\sqrt{\Delta^2+v^2k^2}}J_1(kR)e^{-\sqrt{\Delta^2+v^2k^2}|\tau|}.\nonumber\\
\end{eqnarray}

Using mathematical identity \cite{prudnikov}
\begin{eqnarray}
\int _0^{\infty}x^{\nu+1}\frac{e^{-p\sqrt{x^2+z^2}}}{\sqrt{x^2+z^2}}J_{\nu}(cx)dx=B_{\nu},
\end{eqnarray}
where
\begin{eqnarray}
B_1=c\left(1+z\sqrt{p^2+c^2}\right)\frac{e^{-z\sqrt{p^2+c^2}}}{(p^2+c^2)^{3/2}},\nonumber
\end{eqnarray}
we obtain 
\begin{eqnarray}
\label{gap7}
{\cal G}^{AB}(R;\tau)=\frac{R}{4\pi}\left(1+\sqrt{v^2\tau^2+R^2}\Delta/v\right)\frac{e^{-\sqrt{v^2\tau^2+R^2}\Delta/v}}{(v^2\tau^2+R^2)^{3/2}}.\nonumber\\
\end{eqnarray}
Thus we obtain
\begin{eqnarray}
\label{gap25}
&&\chi^{AB}(R)=-\frac{R^2}{8\pi^2}\int_{-\infty}^{\infty}\left(1+\sqrt{v^2\tau^2+R^2}\Delta/v\right)^2\nonumber\\
&&\frac{e^{-2\sqrt{v^2\tau^2+R^2}\Delta/v}}{(v^2\tau^2+R^2)^3}d\tau.
\end{eqnarray}
In the case $\Delta R/v\gg 1$ we can calculate  integral in Eq. (\ref{gap25}) using the Laplace method, to recover Eq. (\ref{bu}).

\section{Conclusions}

In the end we would like to mention again that though we were considering the case of $T=0$, we have found, that, as it is not infrequently happens, the MGF have  advantages over ZTGF. In particular, using the former one have to operate only with the convergent integrals, in distinction to
what happens when one uses the latter.

\section{Acknowledgments}

Discussions with S. Kettemann,   M. Sherafati, K. Ziegler,
D. Loss, and  K. Szalowski were very illuminating for the author.

\end{document}